\documentclass[11pt]{article}
\usepackage{coling2016}
\usepackage{times}
\usepackage{url}
\usepackage{graphicx}
\usepackage{latexsym}
\usepackage{url}
\usepackage[russian,english]{babel}
\usepackage[utf8]{inputenc}
\usepackage[round]{natbib}
\newcommand{\R}[1]{{\foreignlanguage{russian}{#1}}}

\title{A Visual Representation of\\ Wittgenstein's \textit{Tractatus Logico-Philosophicus}}

\author{
  Anca Bucur\\
  Center of Excellence in Image Study, \\
  Faculty of Letters,\\
  Solomon Marcus Center \\for Computational Linguistics, \\
  University of Bucharest \\
  {\tt anca.m.bucur@gmail.com}\\\And
  Sergiu Nisioi \\ 
  Faculty of Mathematics and \\Computer Science, \\
  Solomon Marcus Center \\for Computational Linguistics, \\
  University of Bucharest \\
  {\tt sergiu.nisioi@gmail.com}  
  }

\date{}

\begin{document}
\maketitle
\begin{abstract}
In this paper we present a data visualization method together with its potential usefulness in 
digital humanities and philosophy of language. We compile a multilingual parallel corpus from different 
versions of Wittgenstein's \textit{Tractatus Logico-Philosophicus}, including the original in German and 
translations into English, Spanish, French, and Russian. Using this corpus, we compute a similarity measure 
between propositions and render a visual network of relations for different languages. 
\end{abstract}

\section{Introduction}
\label{intro}

\blfootnote{
     This work is licensed under a Creative Commons 
     Attribution 4.0 International License.
     License details:
     \url{http://creativecommons.org/licenses/by/4.0/}
}

Data visualization techniques can be essential tools for researchers and scholars in the humanities. In our 
work, we propose one such method that renders concepts and phrases as a network of 
semantic relations. In particular, we focus on a corpus built from different translations of the 
\textit{Logisch-Philosophische Abhandlung} \citep{tractatus} from German into English, French, Italian, 
Russian, and Spanish.

Wittgenstein in his later works states that \textit{meaning is use} \citep{investigations}:
\textit{43. For a large class of cases --though not for all-- in which we employ the word "meaning" it can be 
defined thus: the meaning of a word is its use in the language game. 
And the meaning of a name is sometimes explained by pointing to its bearer.
}

This idea anticipated and influenced later research in semantics, including the
\textit{distributional hypothesis} \citep{harris1954distributional,firth1957synopsis} and more recently, work 
in computational linguistics \citep{lenci2008distributional}. Distributional semantics works on this very 
principle, by making use of data to build semantic structures from the contexts of the 
words. Word embeddings \citep{mikolov2013efficient} are one such example of semantic representation in a 
vector space constructed based on the context in which words occur. 
In our case, we extract a dictionary of concepts by parsing the English sentences and we infer the semantic 
relations between the concepts based on the contexts in which the words appear, thus we construct a 
semantic network by drawing edges between concepts. 

Furthermore, we generalize on this idea to create a visual network of relations between the phrases in 
which the concepts occur.  We have used the multilingual parallel corpora available and created 
networks both for the original and the translated versions. We believe this can be helpful to 
investigate not only the translation from German into other languages, but also how translations into 
English influence translations into Russian, French or Spanish. For example, certain idioms and syntactic 
structures are clearly missing in the original German text, but are visible in both the English and Spanish 
versions.

\section{Dataset}

The general structure of the text has a tree-like shape, the root is divided into 7 propositions, and each 
proposition has its own subdivisions and so on and so forth, in total numbering 526 propositions. A 
\textit{proposition} is the structuring unit from the text and not necessarily propositions 
in a strict linguistic sense. Our corpus contains the 
original German version of the 
text \citep{tractatus} together with translations into 5 different languages: English, Italian, French, 
Russian, 
and Spanish. For English, we have two translations variants, one by \cite{ogden} revised by 
Wittgenstein himself and another one by \cite{pears}.

Since the text has a fixed form structure, it is straight forward to align each translation at the 
proposition level. In addition, we also employ a word-alignment method to create a multilingual parallel 
word-aligned corpus and to be able to inspect how certain concepts are translated into different languages.
The exact size of each version in the corpus\footnote{The dataset is available upon 
request from the authors.} is detailed in Table \ref{dataset}. 
Our corpus contains a relatively small number (526) of aligned examples and alignment methods often fail to 
find the correct pairs between words. To create the word-alignment pairs, we have experimented with different 
alignment strategies including GIZA++ \citep{och2000giza}, fast align \citep{dyer2013simple} and efmaral 
\citep{Ostling2016efmaral}, while the later proved to output the best results in terms of our manual 
evaluation. 

\begin{table}[htb]
\centering
\begin{tabular}{cccc}
\textbf{Language} & \textbf{Translator}         & \textbf{No. of tokens} & \textbf{No. of types} \\
German   & ------           & 18,991        & 4,364         \\
English  & Ogden and Ramsey         & 20,766        & 3,625         \\
English  & Pears \& McGuinness & 21,392        & 3,825         \\
French   & G.G. Granger       & 22,689        & 4,178         \\
Italian  & G.C.M. Colombo     & 18,943        & 4,327         \\
Russian  & M.S. Kozlova       & 10,682        & 4,090         \\
Spanish  & E.T. Galvan         & 13,800        & 3,191        
\end{tabular}
\caption{The size of each corpus in the dataset}
\label{dataset}
\end{table}

The two translations into English share a lot in common, however they are not equivalent, for example, the 
German concept \textit{Sachverhaltes} is translated by \cite{ogden} as \textit{atomic facts} and 
in \cite{pears}'s version the same concept is translated as \textit{states of affairs}. As for the other 
languages, the 
Spanish and Russian translations resemble more the former English version, \textit{Sachverhaltes} being 
translated as \textit{hechos atomicos} and \textit{\R{атомарного факта} (atomarnogo fakta)}, respectively.
In French and Italian, the concept is translated as \textit{états des choses} and \textit{stati di cosi} 
following the \cite{pears} English translation.

\section{Wittgenstein's Network}

\subsection{Tractatus Network}
The \textit{Tractatus Network}\footnote{The \textit{Tractatus Network} is accessible at 
\url{https://tractatus.gitlab.io}} 
is obtained 
from different versions of the text by computing a pair-wise 
similarity measure between propositions. Each proposition is tokenized and each token is stemmed or 
lemmatized. The lemmatizer is available only for English by querying WordNet \citep{fellbaum1998wordnet}, for 
the remaining languages different Snowball stemmers are available in NLTK \citep{bird2009natural}.
Stop words from each proposition are removed before computing the following similarity score:
\begin{equation}
Similarity(p_1,p_2) = \frac{|p_1 \cap p_2|}{max(|p_1|, |p_2|)}
\end{equation}

The similarity score computes the number of 
common tokens between two propositions normalized by the length of the longest proposition, to 
avoid bias for inputs of different lengths. Two propositions are connected by an edge if their similarity 
exceeds the 0.3f threshold. To render the network, we use a browser-based drawing 
library\footnote{\url{http://visjs.org/}}, the lengths of the edges are determined by the similarity value 
and the nodes representing propositions are colored based on the parent proposition (labeled from 1 to 7). 
Furthermore, we added a character n-grams search\footnote{\url{http://fuse.js/}} capability for the network 
that highlights the 
node with the highest similarity to the search string. 

\begin{figure}[hbt]
\centering
\resizebox{0.8\columnwidth}{!}{%
\begin{tabular}{ccc}
\includegraphics[width=3.85cm]{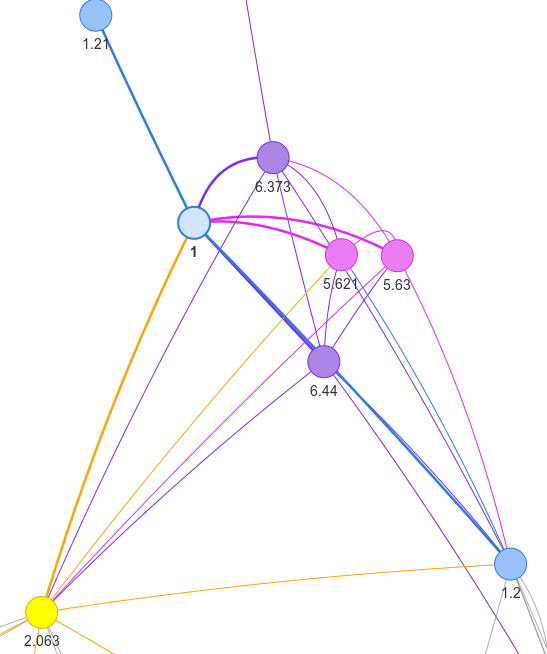} &
\hspace{-0.75em}
\includegraphics[width=3.85cm]{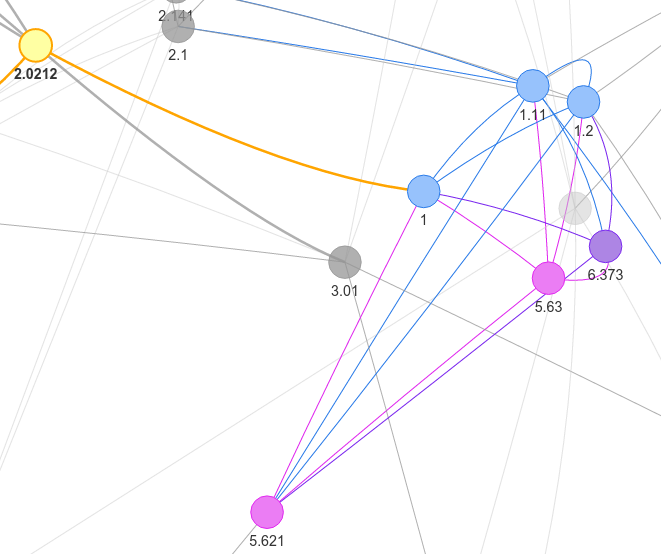} &
\hspace{-0.75em}
\includegraphics[width=3.85cm]{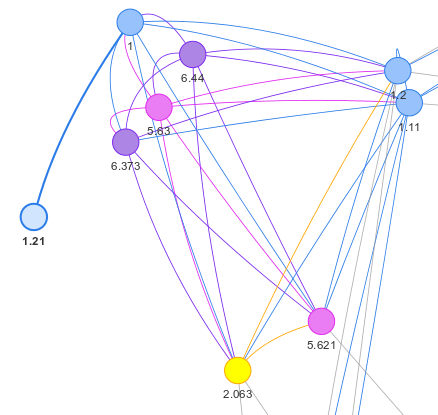} 
\end{tabular}
}
\caption{Two excerpts from the \textit{Tractatus Network}. From left to right we have the German original,
the translations into English by \cite{pears} in the center, and the \cite{ogden} translation on the right. 
Propositions from different groups may resemble each other more than the propositions within the same 
group.}
\label{fig:peamg}
\end{figure}

\begin{figure}[hbt]
\centering
\resizebox{\columnwidth}{!}{%
\begin{tabular}{cccc}
\includegraphics[width=3.85cm]{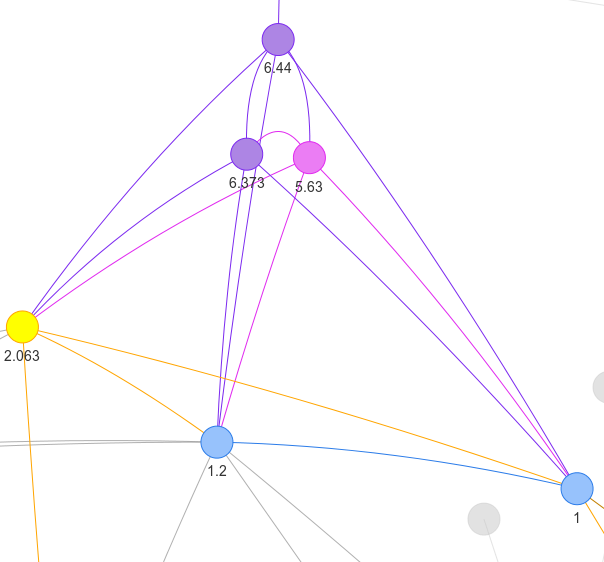} &
 \hspace{-0.75em}
\includegraphics[width=3.85cm]{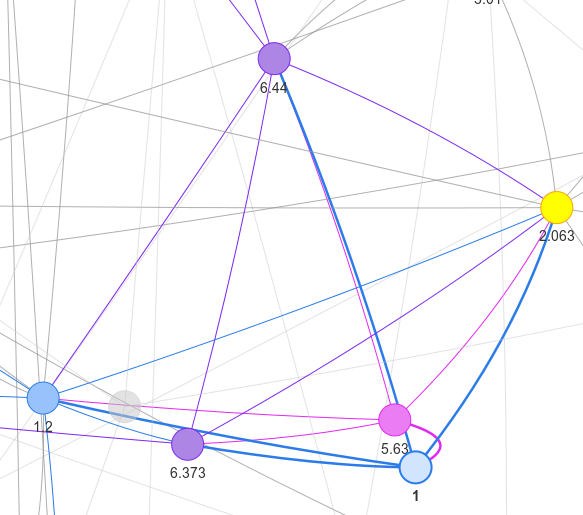} &
 \hspace{-0.75em}
\includegraphics[width=3.15cm]{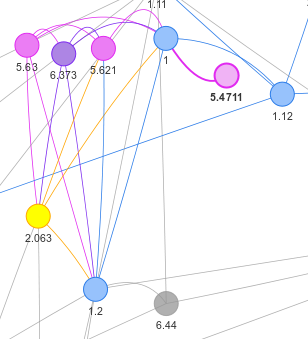} &
 \hspace{-0.75em}
\includegraphics[width=2.8cm]{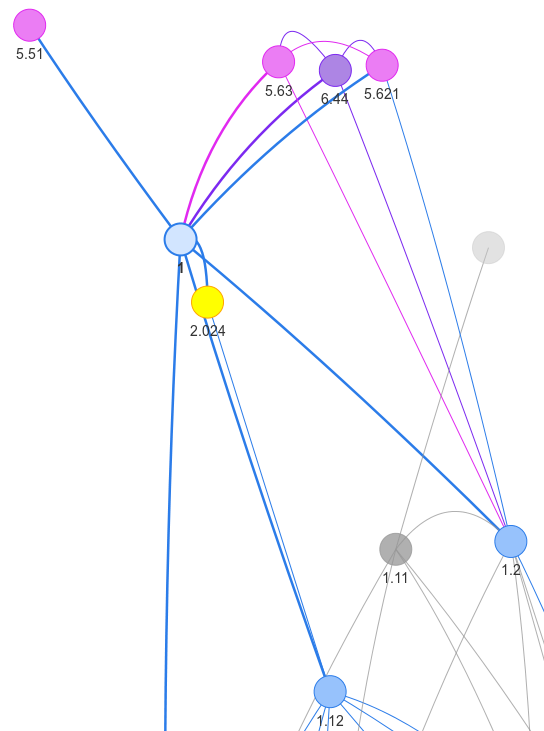} 
\end{tabular}
}
\caption{From left to right: Italian, Spanish, French, and Russian excerpts showing the neighbors of 
proposition 1. Italian and Spanish parts have identical nodes. The French and Russian topologies do not 
resemble the original or any other network.}
\label{fig:romance}
\end{figure}

By analyzing the resulted networks, we can observe that the seven main propositions in the text including 
the sub-divisions are not necessarily hierarchical, at leas not based on the topics addressed, rather the 
\textit{Tractatus} has a rhizomatic structure in which the propositions are entangled and repeatedly make use 
of similar concepts. The excerpts rendered in Figure \ref{fig:peamg} and Figure \ref{fig:romance} bring 
further 
evidence to this observation, as an example the proposition \textit{die gesamte Wirklichkeit ist die 
Welt} meaning \textit{the total reality is the world} appears in almost every version close to the 
propositions in group one in which \textit{die Welt / the world} plays a central role.
In Figure \ref{fig:peamg}, the \cite{pears} English translation  
has a smaller number of relations between propositions, compared to the German counterpart on the left, and 
it also has an additional proposition from group two: \textit{2.0212 In that case we could 
not sketch any picture of the world (true or false)}.
However, in terms of topology, the \cite{ogden} translation resembles almost identically the German version.

On the one hand, looking at the remaining translations, we can observe the Italian and Spanish excerpts 
share the same nodes and comparable topologies with the original German version. On 
the other hand, by looking at the word aligned pairs and the translation of \textit{Sachverhaltes} in 
particular, we may be able trace two separate influences for Spanish and Italian that stem from the different 
English versions of the \textit{Tractatus}.
Last but not least, the French and Russian parts reveal some particularities that cannot be traced to any 
other 
topology from the corpus. 

It is well known that Wittgenstein did not write the propositions in the order they appear in the text and 
our results further evidence this fact by revealing specific clusters of similarity between propositions that 
do not belong to the same group. However, some groups of propositions do appear to be more compact than 
others, e.g. groups 4 and 2 usually have a more compact structure regardless of the language.

\subsection{Concept Network}
The \textit{Concept Network}\footnote{The \textit{Concept Network} is accessible at 
\url{https://wittgenstein-network.gitlab.io}} is created from the main 
concepts/keywords extracted from each proposition in the corpus. For this part, we use only the \cite{ogden} 
translation into English, each proposition is split into sentences and the parse trees are extracted using 
the approach of \cite{honnibal-johnson:2015:EMNLP}. 

\begin{figure}[hbt]
\centering
\resizebox{0.8\columnwidth}{!}{%
\includegraphics[width=3.85cm]{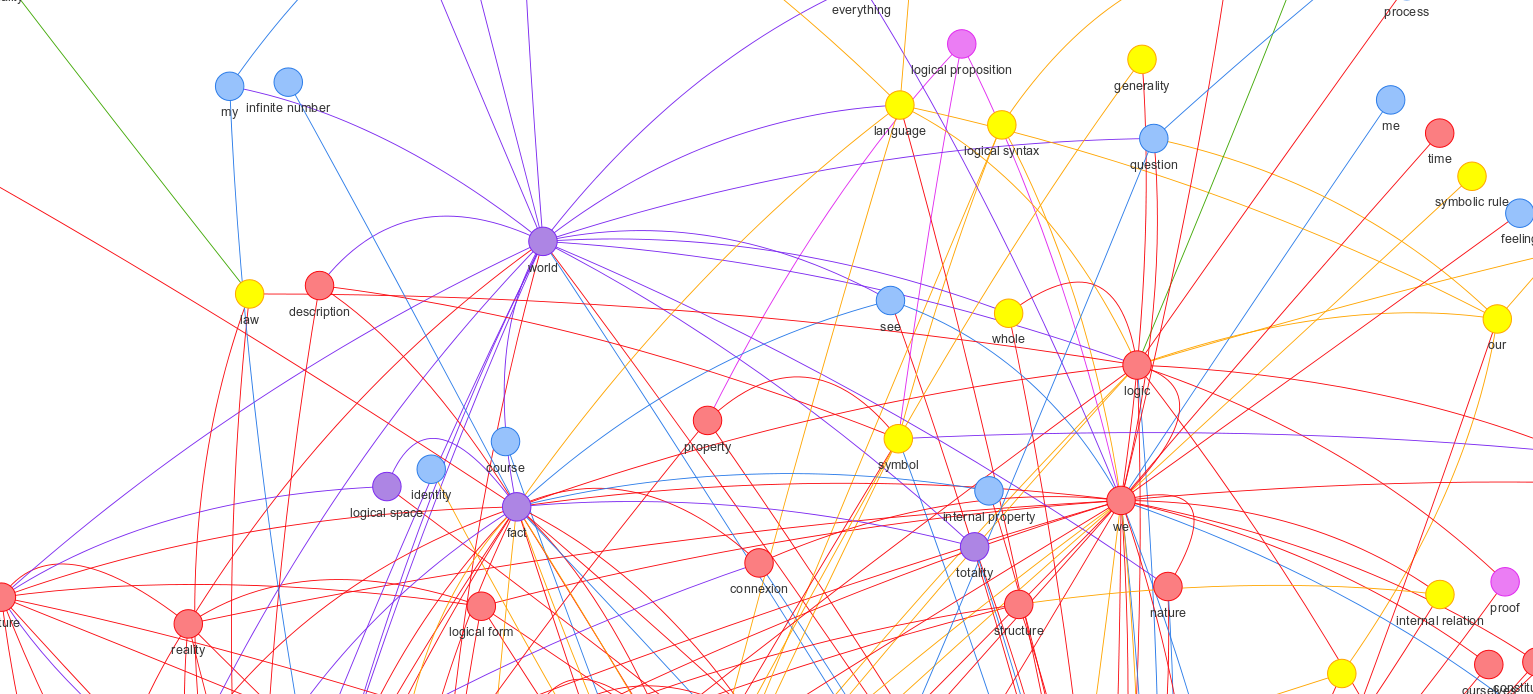}
}
\caption{Excerpt from the concept network. The colors indicate the first group proposition in which the 
concepts appear (from 1 to 7).}
\label{fig:kw}
\end{figure}
The concept list consists of the noun-phrases extracted from the parse trees together with a few personal 
pronouns that appear in the corpus. We manually pruned the 
occurrences having low frequencies and the ones that have been wrongly annotated by the parser. The edges 
between the 
nodes (concepts) are created based on the number of times a concept appears in at least two 
propositions in the same context window, where the window varies depending on how many tokens a concept has. 
Multi word units are allowed to appear in windows of up to ten words, while single token concepts are limited 
to a maximum window of three words.

An excerpt 
from the network is rendered in (Figure \ref{fig:kw}). We noticed that concepts with a high number of edges 
usually occupy a central position in Wittgenstein's philosophy. Words such as: \textit{elementary 
proposition}, \textit{proposition}, \textit{world}, 
\textit{fact}, \textit{form}, \textit{we}, \textit{logic}, \textit{picture},
reveal relations that span across multiple propositions in the text.

\section{Conclusions}
We provide two resources which we believe to be important for scholars and
researchers in digital humanities. The first resource is a compiled, word-aligned corpus extracted from the 
original and translated versions of Wittgenstein's \textit{Tractatus Logico-Philosophicus}. This corpus may 
be used to study the original text or to extract meaningful comparisons from 
translations into other languages. The second resource is a web application that renders semantic networks of 
concepts and propositions from the \textit{Tractatus}. These could be useful to visualize the semantic 
similarities between concepts and to examine the relations between different propositions, to clarify certain 
concepts and to search and explore the actual text, either in German or in translation.
To summarize, therefore, we hope to provide another method of reading Wittgenstein's work.


\section*{Acknowledgements}
This work was supported by a grant of the Romanian National Authority for Scientific Research and Innovation, 
CNCS/CCCDI – UEFISCDI, project number PN-III-P2-2.1-53BG/2016, within PNCDI III

\bibliographystyle{apalike}
\bibliography{bibliografie}

\end{document}